\documentclass[%
 preprint,
 amsmath,amssymb,
 aps, physrev,
]{revtex4-2}

\usepackage{graphicx}
\usepackage{dcolumn}
\usepackage{bm}
\usepackage{hyperref}

\begin{document}


\title{Towards a theory of coupled sociopolitical events-planetary boundaries, crises, policrisis and Earth System syndromes}

\author{Orfeu Bertolami}
\email{orfeu.bertolami@fc.up.pt}
\affiliation{Departamento de F\'{i}sica e Astronomia, Faculdade de Ci\^{e}ncias da Universidade do Porto, Rua do Campo Alegre s/n, 4169-007, Porto, Portugal}
\affiliation{Centro de F\'{i}sica das Universidades do Minho e do Porto, Rua do Campo Alegre s/n, 4169-007, Porto, Portugal}
\author{Ricardo El\'{i}sio}%
\email{rpe@live.com.pt}
\affiliation{Departamento de F\'{i}sica e Astronomia, Faculdade de Ci\^{e}ncias da Universidade do Porto, Rua do Campo Alegre s/n, 4169-007, Porto, Portugal}%


\date{\today}

\begin{abstract}
The impact of the human activities can be evaluated by the Planetary Boundaries (PBs), however, so far it is not clear how to assess the influence of specific sociopolitical events (SPEs) on the Earth System (ES) and at the same measure, without a suitable framework, to gauge how these affect the PBs. In this work, we propose an interacting matrix model that couples SPEs with the PBs and consider the possible evolution scenarios and, in particular, those leading to crises and policrisis. We address specifically the situation where the PBs evolve according to the continuous logistic function, and then consider an exponentially-evolving SPE, which we show to cause a runaway effect on the PBs. We also propose a way to describe, classify, and compare sociopolitical syndromes, that is, a set composed of more than a single polycrisis.
\end{abstract}

\maketitle

\newpage

\section{Anthropocene Misalignments and Earth's System's crises}

We live in a troubled world. The climate crisis, the growth of social inequalities, geopolitical tensions, political turmoil caused by leaders with no ethics, humanity, and preparation to face the challenges of our time are creating a wave of widespread hopelessness and scepticism. There is mounting evidence that an entangled series of crises, or a polycrisis, is lurking at the edge of our time and these will affect every sphere of human activity. Given that in no other instance the effect of human action on the material world has been so overwhelming, it is unquestionable that these crises have a huge impact on the ES, and on the subsystems it comprises at socio-economic and at environmental dimensions \cite{Morin:1999,Lawrence:2024}. As it is well known, the state of the ES can be suitably assessed through the Planetary Boundaries (PB) \cite{Rockstrom:2009,Steffen:2015a}, hence monitoring these variables and how they are directly related to specific sociopolitical events (SPEs) is crucial. 

Indeed, in the transitional period we are living, the Anthropocene, human action has a visible signature on many subsystems of the ES that are nearing their critical point and undergoing a change of regime. Conspicuous examples are the loss of Greenland ice sheet \cite{Lenton:2008}, the continuous degradation of vegetal resilience \cite{Smith:2022}, among many others \cite{Haddad:2015,McKay:2022}. 

These deleterious transitions are driven by human action, which is amplified by the high degree of connectivity of the ES subsystems \cite{Scheffer:2012}, further affecting its general resilience mechanisms \cite{Scheffer:2009}.

As mentioned, the characterization of ES through the PBs allows for an objective evaluation with respect to the Holocene, providing the framework to track its evolution. Recent assessment suggests that six out of the nine PBs show to have evolved well beyond the respective Holocene values \cite{Caeser:2024}. We refer to this divergence of the PBs with respect to what corresponds to a sustainable path of the ES as Anthropocene Misalignments (AMs).

Along Earth's history, the PBs were driven by natural causes: astronomical; geological; and internal dynamics. However, since the second half of last century, human activities have become the overwhelming force and, hence, since then PBs' changes were dominated by the human impact on the ES. These features are at the heart of the thermodynamic description of ES physics through the Landau-Ginsburg phase transition model \cite{BertolamiFrancisco:2018,BertolamiFrancisco:2019}, where the evolution of PBs can be considered as a perturbation of the ES's free energy function at the Holocene \cite{BertolamiFrancisco:2018}.

The bulk of human action can be expressed by the sum of the PBs with respect to their values at the Holocene, $h_i(0)=0$, where the index $i$ runs over the nine PBs and considering their interactions \cite{Barbosa:2020,BertolamiNystrom:2025}:  
\begin{equation}
	H(t) = \sum_{i=1}^9 h_i(t) + \sum_{i,j=1}^9 g_{ij}h_i(t)h_j(t) + \sum_{i,j,k=1}^9 \alpha_{ijk} h_i(t)h_j(t)h_k(t) + ... ,
	\label{Hs_somatorio}
\end{equation}
where $g_{ij} \text{ and } \alpha_{ijk}$ quantify the strength of second and third order interaction terms, respectively. 

Of course, at the Anthropocene, a crucial assumption to establish the evolution of the ES is the time scale of change of function $H$. Indeed, in Ref. \cite{BertolamiFrancisco:2019} it was assumed a linear growth, which was sufficient to conclude that it would lead the ES to evolve, irrespective of the initial conditions \cite{BertolamiFrancisco:2019}, to a Hot-House Earth state \cite{Steffen:2018}. It should be kept in mind that the ES does not possess the same degree of connectivity all the time. In fact, the ES is expected to become more connected in the foreseeable future \cite{Lawrence:2024}. Furthermore, assuming that the PBs follow a discrete logistic map leads to a richer set of ES trajectories, featuring bifurcations and chaotic behaviour \cite{Bernardini:2025}. In Ref. \cite{BertolamiNystrom:2025} it was shown that the ES resilience can, in physical terms, be associated with the existence of metastable states and with the dynamical dissipation of energy, which tend to halt, or at least slow down the evolution of the ES towards the Hot-House Earth. 

The Landau-Ginsburg model described above has been devised to account for the ES's evolution, given a prototypical evolution the impact human action exerts on the PBs. However, in order to analyse the impact of specific SPEs on the ES, their capability to engender a crisis or crises, and trigger the onset of a polycrisis, it is necessary to establish how SPEs are affected by the evolution of the PBs and, on the other way around, how the SPEs back-react on the PBs. In this work, we propose a simple matrix interaction model that couples the PBs to the SPEs. 

We assume that the SPEs are described by a vector $\vec{E}(\vec{r},t)$, and the PBs by a vector $\vec{h}(\vec{r},t)$, where we shall assume for simplicity, given the number of PBs, that the SPE vectors also  belong to the same 9-dimensional vector space. For simplicity, we shall drop the space dependence of these vectors. The concrete assignments for each of the SPEs will be described below. From this setup, we can write a generic interaction term $R_{ij}E_ih_j$, where the features of matrix $R_{ij}$, spanning $i,j= 1, 2, ..., 9$, will be promptly specified. 

This work is organized as follows. In section 2, we shall set up the framework for an interaction model of vectors $\vec{E}(t)$ and $\vec{h}(t)$ and the Lagrangian/Hamiltonian formalism to obtain their equations of motion. These will be then solved assuming that the PBs evolve as the logistic function. Complementarily, we shall also show that an exponentially evolving SPE leads to an exponential runaway of the PBs. In section 3, we consider gathering of SPE vectors whose components evolve as crises, that is, have a positive derivative in time. We construct a polycrisis vector out of the components of SPEs with at least two of these components, and introduce the concept of "Syndromes", built out of the vector product of polycrisis vectors. We also sketch how resilience can be introduced into the discussion. Finally, in section 4 we present our conclusions.   

\section{ A matrix model to couple Planetary Boundaries to Socio-Political Events}

As discussed above, a descriptive and quantitative model of the ES's crises and their evolution can be achieved through a matrix model, where the SPEs vector interacts with the PBs vector. The SPEs are contingent and somewhat arbitrary, but can be considered to correspond to a class of events such as: "Economic Growth", "Harmful Technological Developments", "Bellicose Actions", "Social Inequalities", "Poverty and Famine", "Pandemics", "Ecological Disasters", "Unemployment" and "Disruptive Political Developments". Likewise, the "strength" of initial values of these components is arbitrarily set to lie in the range from, say, 0 to 5, where the lowest limit means no cause/effect, while the upper limit means the strongest possible onset for the SPE. As stated above, the PBs are assumed to vanish at the Holocene and to acquire contributions, set by the current magnitude of the human impact on each specific PB.                                     

Obviously, SPEs have their own dynamics and are most often driven and fueled by other SPEs. These features will be developed elsewhere. In here, we are concerned with the dynamics that connect the SPEs and the ES through their impact on the PBs. Our approach and definition of the SPEs vector is inspired on the idea that some socioeconomic and political developments lead to degradation of the ES, or, as proposed in Ref \cite{Jorgensen:2024}, to Anthropocene Traps. As mentioned, we refer to the growth of each SPE as a crisis and its impact on the PBs as Anthropocene misalignments.

There is a broad range of complex systems that allow for a description in terms of matrix models. Examples in condensed matter physics include models of ferromagnetism, quantum Hall effect, topological insulators (see e.g. Refs. \cite{Ising:1925,Halperin:1982,Haldane:1983}). In string theory, matrix models provide a non-perturbative approach of strings theory in low-dimensional spacetimes \cite{BFSS:1997}. In the present work we shall consider these systems as a model to describe the complex interaction of vectors $\vec{E}(t)$ and $\vec{h}(t)$. That is to say, we assume that these vectors interact via a matrix term $R_{ij}$, whose entries we conveniently choose to lie in the range $-1 \leq R_{ij} \leq 1$. This choice of values will determine the degree of alignment or misalignment between the PBs and SPEs vectors. 
The interaction matrix can be a dynamical element of our model, also exhibiting spacial and/or time dependence, which we shall ignore here. This is clearly a simplification, but a natural one, given that most relevant features of the anthropogenic activities can already be captured by the evolution of the vectors $\vec{E}(t)$ and $\vec{h}(t)$, with no need of further introducing more complexity through a more elaborate version of the interaction matrix, $R_{ij}$. 

Given that we are interested in obtaining evolution equations for the SPEs from a set of PBs and vice versa, we consider then the Lagrangian/Hamiltonian framework, whose dynamics can be described, to start with, by the Lagrangian function: 
\begin{equation}
	\label{Lagrangeano_PB}
	L(\vec h, \vec E, \dot{\vec h}, \dot{\vec E};t) =  \frac{1}{2} \mu \sum_{i=1}^9 \dot{h_i}^2 + \frac{1}{2} \nu \sum_{i=1}^9 \dot{E_i}^2 - \sum_{i,j=1}^9 E_iR_{ij}h_j,
\end{equation}
where $\mu$ and $\nu$ are arbitrary phenomenological constants and the dots denote time derivatives.

From the Lagrangian function we can obtain the canonical conjugate momenta:
\begin{align*}
	p_{hi} = \frac{d L}{d \dot h_i} = \mu \dot h_i, \\
	p_{Ei} = \frac{d L}{d \dot E_i}  = \nu \dot E_i,
\end{align*}
which enables to get the Hamiltonian function of the model:
\begin{equation}
    \label{Hamiltoniana_PB}
    H(\vec h, \vec E, \vec p_h, \vec p_E;t) = \sum_{i=1}^9 \frac{p_{hi}^2}{2\mu} + \sum_{i=1}^9 \frac{p_{Ei}^2}{2\nu} + \sum_{i,j=1}^9R_{ij}E_ih_j.
\end{equation}

Straightforward manipulation of Eq. \eqref{Hamiltoniana_PB} results in the dynamical equations of the model:

\begin{align}
	\label{Momento_b_ponto}
	\dot p_{hi} &= - \sum_{j=1}^{9} E_j R_{ji}, \\
	\label{b_ponto}
	\dot h_i &= \frac{p_{hi}}{\mu}, \\
	\label{Momento_a_ponto}
	\dot p_{Ei} &= - \sum_{j=1}^{9} R_{ij}h_i, \\
	\label{a_ponto}
	\dot E_i &= \frac{p_{Ei}}{\nu}.
\end{align}

From this set of equations, the SPE's evolution is coupled to the PBs and vice-versa. 

\begin{figure}[h!]
	\centering
	\includegraphics[scale=0.75]{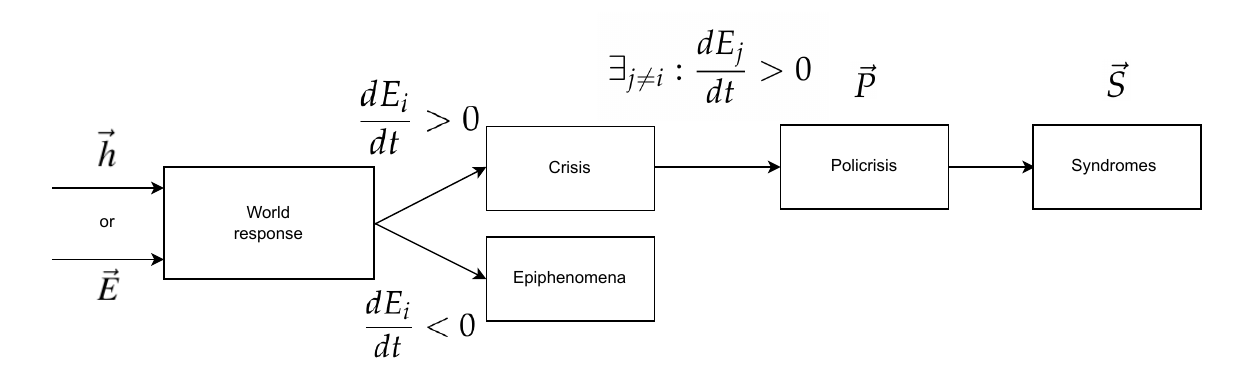}
	\caption{Pictoric scheme of the model's structure.}
	\label{Esquema_modelo}
\end{figure}

We sketch in \autoref{Esquema_modelo}, the structure of our interacting model. 

The symmetry of the interaction term with respect to $\vec E (t)\text{ and } \vec h(t)$ vectors and obvious rotational invariance allow for considering two complementary approaches. We can consider the PBs vector as an input and obtain the SPE vector, or the other way around. These two types of solutions will be considered in the next subsection. 

It should be pointed out that, as indicated in \autoref{Esquema_modelo}, depending on the evolution of the components $E_i(t)$ of the vector $\vec{E}(t)$, one expects a crisis if $dE_i(t)/dt > 0$, or an epiphenomenon if $dE_i(t)/dt < 0$, once the evolution of vector $\vec{h}(t)$ is specified. A polycrisis \cite{Lawrence:2024} arises once at least two components of the SPEs vector have positive derivatives in time. From these components we can build a polycrisis vector, $\vec P (t)$. As we shall discuss later on, a kth-order syndrome will be defined out of the external product of $k$ polycrisis vectors. Of course, each crisis is fed by the interaction with the PBs, that is, with the ES. 

It is worth pointing out that our definition of a crisis as a SPEs growing function in time is consistent with the one arising from the common understanding in social sciences. Indeed, formally, a crisis is defined as a sudden event that causes local or global harmful effects on a large portion of the human population \cite{Lawrence:2024}. A set of crises can get entangled and overlap \cite{Morin:1999,Lawrence:2024} due to the high degree of connectivity of the human activities, of the ES itself and on the effect the former has on the latter.

In the next subsection, we shall examine solutions of our model assuming that the PBs evolve as a logistic function and then consider the model's response to a SPE that evolves exponentially.

\subsection{The logistic function for the evolution of the Planetary Boundaries}

The model proposed above allows one to obtain the evolution of a SPE once the behaviour of the PBs is known. In Ref. \cite{BertolamiFrancisco:2019} it was assumed that, as a whole, $H(t)$ had a linear growth in time, which is a fair choice for a short time interval. It was argued, however, that a more suitable behaviour for the PBs was to consider the discrete logistic map \cite{Bernardini:2025}. In here, we shall consider that a sensible growth assumption for a longer time period is the continuous logistic function \cite{Verhulst:1838}, as it captures both the accelerated initial exponential "Malthusian" growth, as well as the inevitable decay due to the finiteness of resources and the impact that human activities have on the ES. Thus, we consider that the growth rate of each PB is given by \cite{Verhulst:1838}:
\begin{equation}
	\label{Hipotese_logistica}
	\dot h_i(t) = \lambda_i h_i(t) \left[1 - K_ih_i(t)\right],
\end{equation}
where $\lambda_i > 0$ is the growth factor for the $i$th PB and the inverse of $K_i$ sets the upper bound value for $h_i$, once $K_i > 0$, which will be assumed throughout the discussion. This differential equation admits the well known solution \cite{Verhulst:1838}:
\begin{equation}
	\label{Solução_logistica}
      h_i(t) = \frac{e^{\lambda_it}}{C_i + K_ie^{\lambda_it}},
\end{equation}
being $C_i$ a positive integration constant. 

Introducing Eq. \eqref{Momento_a_ponto} in the time derivative of Eq. \eqref{a_ponto} and setting, without any loss of generality, that $\mu=\nu=1$, yields: 
\begin{equation}
	\label{Rel_app_b}
	\ddot E_i(t) = -\sum_{j=1}^{9} R_{ij}\frac{e^{\lambda_jt}}{C_j + K_je^{\lambda_jt}}.
\end{equation}

One sees that due to the interaction matrix a rich lore of solutions can arise from the sum over the PBs, weighted differently for all components of the SPEs vector. Performing the first integration step yields:
\begin{equation}
\begin{aligned}
    \dot E_i(t) - \dot E_{i0} =& -\sum_{j=1}^9 \frac{R_{ij}}{ K_j\lambda_j}\left[ \ln \left| C_j + K_j e^{\lambda_jt} \right| - \ln\left|C_j+K_j\right| \right]\\
    =&  -\sum_{j=1}^9 \frac{R_{ij}}{ K_j\lambda_j} \left[ \ln \left| C_je^{-\lambda_jt} + K_j \right| + \lambda_j t - \ln\left|C_j+K_j\right| \right].
    \end{aligned}
    \label{E_ponto}
\end{equation}

Further integration leads to the expression,
 
\begin{equation}
\begin{aligned}
    E_i(t) - \dot {\tilde{E}}_{i0}t-E_{i0} =& -\sum_{j=1}^9 \frac{R_{ij}}{ K_j\lambda_j} \int_{0}^{t} \left[ \ln \left| C_je^{-\lambda_jt^\prime} + K_j \right| + \lambda_jt^\prime \right] dt^\prime \\
    =&  -\sum_{j=1}^9 \frac{R_{ij}}{K_j\lambda_j} \left[ \frac{1}{\lambda_j} Li_2\left( -\frac{C_j}{K_j}e^{-\lambda_jt} \right) - \frac{1}{\lambda_j} Li_2\left( -\frac{C_j}{K_j} \right) + (\ln|K_j|) t + \frac{\lambda_j}{2} t^2\right],
    \end{aligned}
    \label{E_geral}
\end{equation}
where the constant $\dot E_{i0}$ was adjusted and the dilogarithm function is defined as follow:
\begin{equation}
	\label{Dilogaritmo}
	Li_2(z) \equiv -\int_0^z \frac{\ln(1-z^\prime)}{z^\prime} dz^\prime,
\end{equation}
for $z \in \mathbb{C} \setminus_{[1,+\infty[}$. As expected, a general SPE solution is given in terms of two arbitrary constants, $\{ E_{i0},\dot E_{i0}\}$. It is worth remarking that Eq. \eqref{E_geral}, thanks to the interaction matrix, allows for evaluating the SPE component at any given moment in time. In order to exploit this predictability, it is useful to obtain a closed expression for the above solution. Refraining from making unjustified assumptions about the ratio $C_j/K_j$, the most general approach is to consider before integration the expansion
\begin{equation}
    \label{serie_aprox}
    f(z) \equiv \frac{\ln{}(1-z)}{z}=-\sum_{q=0}^\infty \frac{z^{q}}{q+1}\text{,  }|z|<1.
\end{equation}
To proceed one considers the Padé approximation technique \cite{Baker:1996,Wanner:1978} given its robustness and versatility. In order to track the accuracy of the approximation, one considers the ratio between polynomials, $[L/M](z)$, that will represent the function Eq. \eqref{serie_aprox}:
\begin{equation}
    \label{razao_aproximacao}
    s(z) \equiv \frac{[L/M](z)}{f(z)}=\frac{\sum_{q=0}^La_qz^q}{1+\sum_{k=1}^Mb_kz^k}\frac{z}{\log\left( 1-z \right)}, 
\end{equation}
which, in its turn, will guide the choice of polynomials to get the profile function one desires. Eq. \eqref{razao_aproximacao} gives origin to what is called an order star \cite{Baker:1996,Wanner:1978} in the complex plane. The best approximation is obtained, when, in the region of interest of the "star", $s(z) \approx 1$. For $z \approx 0$, the original function is well represented by a MacLaurin series. The Padé approximants will be constructed with this series' coefficients, which are sufficiently reliable near the origin. Furthermore, to ensure the best fit for large $z$, $[L/M](z)$ must behave asymptotically as $z^{-1}$ so to match the behaviour of $f(z\rightarrow-\infty)$. Therefore, it is necessary that the degree of the polynomial at the numerator, $L$, and the degree of the polynomial at the denominator, $M$, satisfy the condition $L=M-1$. In fact, the order of the polynomials can be decided by the required accuracy one aims to achieve. The approximation's maximum error is tied to the ratio $r\equiv C_j/K_j$ and decays with time. So, assuming that $r \lesssim 10$, the following $5^\text{th}$ order approximation constrains the relative error below $3\%$:
\begin{equation}
\label{aproximante23}
	f(z) \simeq \frac{11z^2-60z+60}{3z^3-36z^2+90z-60}.
\end{equation}

This analytic approximation allows for a suitable integration and evaluation of the dilogarithmic function
\begin{equation}
	\label{Dilogaritmo_aproximado}
	\begin{aligned}  
            Li_2\left( z \right)\simeq& -\Bigg[\frac{25}{6\sqrt{15}}\ln\left| \frac{z-\sqrt{15}-5}{z+\sqrt{15}-5}\right| + \frac{25}{18}\ln\left| z^2-10z+10 \right| + \frac{8}{9}\ln\left|z-2\right|\Bigg]_0^{-\frac{C_j}{K_j}e^{-\lambda_jt}}.
	\end{aligned}
\end{equation}

After substitution of the integration limits and considering the constant dilogarithm function in Eq. \eqref{E_geral}, one obtains:
\begin{equation}
    \begin{aligned}
        Li_2\left( -\frac{C_j}{K_j}e^{-\lambda_jt} \right) -  Li_2\left( -\frac{C_j}{K_j} \right) \simeq& -\frac{25}{6}\Bigg[ \frac{1}{\sqrt{15}}\ln\left| \frac{C_je^{-\lambda_jt}+K_j\sqrt{15}+5K_j}{C_je^{-\lambda_jt}-K_j\sqrt{15}+5K_j}\cdot\frac{C_j-K_j\sqrt{15}+5K_j}{C_j+K_j\sqrt{15}+5K_j} \right| \\
        +& \frac{1}{3}\ln\left| \frac{C_j^2e^{-2\lambda_jt}+10C_jK_je^{-\lambda_jt}+10K^2}{C_j^2+10C_jK_j+10K_j^2} \right|\Bigg] \\
        -&\frac{8}{9}\ln\left| \frac{2K_j+C_je^{-\lambda_jt}}{2K_j+C_j} \right|.
    \end{aligned}
    \label{Dilogaritmo_final}
\end{equation}

From now on, the constants inside the sums will be considered through suitable redefinitions of $E_{i0}$ so to obtain neater expressions. 

Let us consider the two most relevant limits for analysing the evolution of a SPE. For $t\ll \lambda_j^{-1}$, the contribution of the dilogarithms is negligible and the behaviour of $E_i(t)$ is controlled by the linear term: 
\begin{equation}
    \label{E_assintotico_t_pequeno}
    \begin{aligned}
    E_i(t) \sim \dot{\tilde{E}}_{i0}t + \tilde{E}_{i0} -\sum_{j=1}^9 \frac{R_{ij}}{ K_j\lambda_j^2} \Bigg\{& (\ln|K_j|)\lambda_jt + \frac{\lambda_j^2}{2} t^2 \\
    -& \Bigg( \frac{25}{6}\Bigg[ \frac{1}{\sqrt{15}}\ln\left| \frac{C_j(1-\lambda_jt)+K_j\sqrt{15}+5K_j}{C_j(1-\lambda_jt)+K_j\sqrt{15}-5K_j} \right| \\
    +& \frac{1}{3}\ln\left| -(2C_j^2 + 10C_jK_j)\lambda_jt + C_j^2+10C_jK_j+10K_j^2 \right|\Bigg] \\
    +&\frac{8}{9}\ln\left|C_j(1-\lambda_jt) +2K_j \right| \Bigg) \Bigg\}.
    \end{aligned}
\end{equation}

On the other hand, for $t\gg\lambda_j^{-1}$, the asymptotic limit is controlled by terms of the same order as in the previous expansion since the dilogarithmic contribution becomes constant:
\begin{equation}
    \begin{aligned}
        E_i(t) \sim \dot {\tilde{E}}_{i0}t + \tilde{E}_{i0} -\sum_{j=1}^9 \frac{R_{ij}}{ K_j\lambda_j^2} \Bigg\{ (\ln|K_j|)\lambda_jt + \frac{\lambda_j^2}{2} t^2& - \Bigg( \frac{25}{6}\Bigg[ \frac{1}{\sqrt{15}}\ln\left| \frac{\sqrt{15}+5}{\sqrt{15}-5} \right| \\
        +& \frac{1}{3}\ln\left|10K_j^2 \right|\Bigg] + \frac{8}{9}\ln\left(2K_j \right) \Bigg) \Bigg\}, \\
    \end{aligned}
    \label{E_assintotico_t_grande}
\end{equation}
that is
\begin{equation}
    \begin{aligned}
        E_i(t) \sim \dot{\tilde{ E}}_{i0}t + \tilde{E}_{i0} -\sum_{j=1}^9 \frac{R_{ij}}{ K_j\lambda_j^2} \Bigg\{& (\ln|K_j|)\lambda_jt + \frac{\lambda_j^2}{2} t^2 \Bigg\},
    \end{aligned}
    \label{E_assintotico_t_grande1}
\end{equation}
where the pair of integration constants was suitably redefined as:
\begin{equation}
    \label{Redefinicao_constantes}
    \begin{aligned}
        \dot{\tilde{ E}}_{i0} &\equiv \dot E_{i0} + \sum_{j=1}^9\frac{R_{ij}}{K_j\lambda_j}\ln\left| C_j+K_j \right| , \\
        \tilde{E}_{i0} &\equiv E_{i0} + \sum_{j=1}^9\frac{R_{ij}}{K_j\lambda_j^2} \Bigg\{ \frac{25}{6\sqrt{15}}\ln\left| \frac{C_j+5K_j-K_j\sqrt{15}}{C_j+5K_j+K_j\sqrt{15}} \right| \\
        &- \frac{25}{15}\ln \left| C_j^2+10C_jK_j+10K_j^2\right| - \frac{8}{9}\ln\left| C_j+2K_j\right|\Bigg\}.    
        \end{aligned}
\end{equation}

Thus, one sees that the logistic assumption for the PBs implies that, for large times, the SPEs evolve quadratically, as depicted in Eq. \eqref{E_assintotico_t_grande}. As expected, for small time, SPEs evolve linearly as shown in Eq. \eqref{E_assintotico_t_pequeno}. 

In fact, the behaviour obtained above is dictated by the logistic function, Eq. \eqref{Hipotese_logistica}, for the PBs. If we had considered instead the "Malthusian" exponential behaviour, swiftly obtained taking $K_i\rightarrow0$, we would get exponentially growing SPE components. 

Notice that for describing an evolving crisis that tends to become worse, that is $dE_i(t)/dt>0$, it is required that $-1 \le R_{ij} < 0$.  

Given the symmetry of our model, we can consider, in opposition to what was examined above,  the impact of the SPEs on the PBs. In particular, let us consider a full-blown crisis with an exponential growth. That is, we consider: 
\begin{equation}
    \label{SPE_exponencial}
    E_i(t)=A_ie^{B_it},
\end{equation}
where $A_i \text{ and } B_i$ are constants. This leads to the $h_i(t)$ term, the equation of motion (see Eqs. \eqref{Momento_b_ponto} and \eqref{b_ponto}):
\begin{equation}
    \begin{aligned}
        \ddot h_i(t) =& -\sum_{j=1}^9\frac{R_{ji}}{\mu}E_j(t),
    \end{aligned}
\end{equation}
whose integration leads, as expected, to the result
\begin{equation}
    h_i(t) - \dot h_{i0}t - h_{i0} = - \sum_{j=1}^9 R_{ji}\frac{A_je^{B_jt}}{B_j^2}.
    \label{SPE_evol_expon}
\end{equation}
Thus, an exponential crisis leads to an exponential runaway of the PBs. This type of behaviour has dire effects on the ES \cite{Rockstrom:2009,Steffen:2015a}. Notice, that the impact of the exponential growth in Eq. \eqref{SPE_exponencial} is visible even for a slowly growing exponential, as the constant $B_j$ appears in the denominator of solution Eq. \eqref{SPE_evol_expon}.

Given the high degree of connectivity of the ES and of the SPEs, it is natural to expect that a crisis gives origin to a set of ensued crises, a polycrisis. In the next section, we shall consider this situation and introduce the concept of Syndromes in order to better characterise how the SPEs resonate on the ES. 

\section{From Socio-Political Events to Syndromes}

Having shown how the SPEs-PBs coupled model works, we address now the issue of how sets of crises can, for instance, be compared with each other. This opens the possibility of capturing more objectively the concept of a polycrisis. Needless to say that this characterisation might be relevant to draw concrete policies to tackle and mitigate them whenever possible. This characterisation is achieved through the concept of Syndromes, that we address in the following.

As previously discussed, at their onset, SPEs are arbitrarily and qualitatively gauged to range from 0 to 5. However, through the way they couple to the PBs via the matrix $R_{ij}$ one can track their evolution. For a specific set of SPEs, labelled with an upper index $l$, $\vec{E}^{(l)}(t)$, a crisis ensues, fed by the PBs, if $dE_i^{(l)}(t)/dt >0$. Given the dynamic coupling to the PBs, crises arising at different times evolve differently, hence it is natural to distinguish different sets of SPEs, even though in the same qualitative class, and consider their gathered effect. The vector nature of the SPEs allows one to consider the cross product of two or more SPEs vectors as a measure of the resulting generalised volume \footnote{Given the SO(n) symmetry of the interaction, the emerging space structure of the Syndromes correspond to the one of k-forms in $l$ dimensions \cite{Schutz:1980}.}. 

Clearly, a single SPE vector, with at least two components that are growing functions of time, correspond to a single polycrisis. These components may be collected into a polycrisis vector $\vec P^{(1)}_l(t)$ with $l$ components, where $l \le 9$ refers to the number of crises contained in a given set of SPEs $\vec E^{(1)}$. In this vector subspace it is quite natural to define a kth-order Syndrome, from $k$ intermediate polycrisis vectors in a certain historical moment,
\begin{equation}
	\label{syndromes_def}
	S^{(k)}_l(t) = \underbrace{\vec P_l ^{(1)}(t) \times \vec P_l^{(2)}(t) \times ... \times \vec P_l ^{(k)}(t)}_{\text{$k$ times}},
\end{equation}
given that $k \le l$. Note that each of the $k$ polycrisis vectors refers to sets of SPEs that are independent and do not necessarily emerge simultaneously. By the means of this construction one is able to collect all SPE components that depict a crisis behaviour in suitable sets of vectors $\vec P$, while simultaneously considering their intertwining through Syndromes of order $k$.

As discussed in the previous section, it is obvious that a full-blown exponentially evolving crisis will ensue an exponential deterioration of the PBs, evinced in Eq. \eqref{SPE_evol_expon}, implying in an obvious Anthropocene misalignment. The connection between these misalignments and the Anthropocene traps \cite{Jorgensen:2024} is still to be worked out in detail, but is evident given that the former imply necessarily in a poorer performance of the ES in what concerns its stability and sustainability. Moreover, the "stronger" the AM among the SPEs and PBs vectors, the more the ES is pulled towards degradation, resembling the Anthropocene Traps, defined in Ref. \cite{Jorgensen:2024}.

An obvious advantage of our description via Syndromes is that it encapsulates all the desired interactions amongst the SPEs. Furthermore, it allows for an objective comparison between Syndromes through the volume they yield, suggesting that a fair comparison between different sets of polycrisis can only be achieved if they are of the same kth-order Syndrome class.

Before closing this discussion, let us point out that our modelling admits various ways to introduce resilience into the description. As proposed in Ref. \cite{BertolamiNystrom:2025}, in the thermodynamic model of the ES, resilience can be considered in terms of the presence of metastable states and energy dissipation. The latter is an intrinsic feature of any physical system, the former has to be introduced via restoring ecossystems and properly designed adaptation, mitigation and geoengineering strategies. In the coupled SPEs-PBs model discussed here, resilience can be introduced, for instance, considering that the parameters of the logistic evolution of the PBs can be modified, that is:   
\begin{equation}
\begin{aligned}
	\lambda_j \rightarrow \lambda_j (t) , \\
	K_j \rightarrow K_j(t),
\end{aligned}
\label{Time dep}
\end{equation}
according to some functional dependence that can be established empirically. Of course, this dependence introduces complexity into our description, but it has a predictable effect on the evolution of the SPEs via an equation equivalent to Eq. \eqref{Rel_app_b} with a suitable evolution for $h_i(t)$ compatible with the time dependence expressed through Eqs. \eqref{Time dep}. 

Resilience features are required if crises and syndromes are to be more promptly attenuated. Qualitatively, one aims to relax the connectivity of the ES so to ensure that crises do not affect significantly the PBs, or in other words, one aims to effectively switch off the interaction matrix, that is $|R_{ij}|\rightarrow 0$, through a set of suitably designed strategies. These would transform crises and syndromes into harmless epiphenomena, stabilized by the ES's internal processes.

\section{Discussion and Conclusions}

In this work we have considered a model where SPEs and PBs are coupled by a matrix interaction, likewise some complex physical model in condensed matter and string theory. This is achieved via the assumption that both, SPEs and PBs, are elements of a 9-dimensional vector space, whose dynamics in the phase space arise from the Lagrangian/Hamiltonian formalism built out of these vectors. The resulting equations of motion correspond to trajectories on the phase space once a law for the evolution of one of these vectors is imposed. We have shown that the matrix interaction allows for obtaining short term and asymptotically meaningful solutions for the SPEs vectors, once the logistic function evolution is assumed for the PBs. An exponential evolution of any of the vectors leads to an exponential evolution of the other vector. 

We identify a crisis when a component of the SPE vector evolves such as $d E_i(t)/dt > 0$ and a polycrisis if at least two components of a SPE vector have positive derivatives. From a set of $k$ polycrisis, we build up the concept of a kth-order syndrome via the vector product of different polycrisis vectors (cf. Eq. \eqref{syndromes_def}). Syndromes are particularly harmful AM, given the degradation they cause on the PBs, and they can be objectively compared with each other provided they lie in the same order. 

It is worth pointing out that tracking the PBs' evolution once a SPE unfolds allows for the possibility of drawing more effective adaptative and mitigatory actions, envisaging to avoid further deterioration of the working conditions of the ES.  

\bibliography{apssamp}

\begin{thebibliography}{26}%
\makeatletter
\providecommand \@ifxundefined [1]{%
 \@ifx{#1\undefined}
}%
\providecommand \@ifnum [1]{%
 \ifnum #1\expandafter \@firstoftwo
 \else \expandafter \@secondoftwo
 \fi
}%
\providecommand \@ifx [1]{%
 \ifx #1\expandafter \@firstoftwo
 \else \expandafter \@secondoftwo
 \fi
}%
\providecommand \natexlab [1]{#1}%
\providecommand \enquote  [1]{``#1''}%
\providecommand \bibnamefont  [1]{#1}%
\providecommand \bibfnamefont [1]{#1}%
\providecommand \citenamefont [1]{#1}%
\providecommand \href@noop [0]{\@secondoftwo}%
\providecommand \href [0]{\begingroup \@sanitize@url \@href}%
\providecommand \@href[1]{\@@startlink{#1}\@@href}%
\providecommand \@@href[1]{\endgroup#1\@@endlink}%
\providecommand \@sanitize@url [0]{\catcode `\\12\catcode `\$12\catcode
  `\&12\catcode `\#12\catcode `\^12\catcode `\_12\catcode `\%12\relax}%
\providecommand \@@startlink[1]{}%
\providecommand \@@endlink[0]{}%
\providecommand \url  [0]{\begingroup\@sanitize@url \@url }%
\providecommand \@url [1]{\endgroup\@href {#1}{\urlprefix }}%
\providecommand \urlprefix  [0]{URL }%
\providecommand \Eprint [0]{\href }%
\providecommand \doibase [0]{https://doi.org/}%
\providecommand \selectlanguage [0]{\@gobble}%
\providecommand \bibinfo  [0]{\@secondoftwo}%
\providecommand \bibfield  [0]{\@secondoftwo}%
\providecommand \translation [1]{[#1]}%
\providecommand \BibitemOpen [0]{}%
\providecommand \bibitemStop [0]{}%
\providecommand \bibitemNoStop [0]{.\EOS\space}%
\providecommand \EOS [0]{\spacefactor3000\relax}%
\providecommand \BibitemShut  [1]{\csname bibitem#1\endcsname}%
\let\auto@bib@innerbib\@empty
\bibitem [{\citenamefont {Morin}\ and\ \citenamefont
  {Kern}(1999)}]{Morin:1999}%
  \BibitemOpen
  \bibfield  {author} {\bibinfo {author} {\bibfnamefont {E.}~\bibnamefont
  {Morin}}\ and\ \bibinfo {author} {\bibfnamefont {A.}~\bibnamefont {Kern}},\
  }\bibinfo {title} {Homeland earth: A manifesto for the new millennium}\
  (\bibinfo  {publisher} {Hampton Press, Inc.},\ \bibinfo {year} {1999})\ pp.\
  \bibinfo {pages} {72--75}\BibitemShut {NoStop}%
\bibitem [{\citenamefont {Lawrence}\ \emph {et~al.}(2024)\citenamefont
  {Lawrence}, \citenamefont {Homer-Dixon}, \citenamefont {Janzwood},
  \citenamefont {Rockstr{\"o}m}, \citenamefont {Renn},\ and\ \citenamefont
  {Donges}}]{Lawrence:2024}%
  \BibitemOpen
  \bibfield  {author} {\bibinfo {author} {\bibfnamefont {M.}~\bibnamefont
  {Lawrence}}, \bibinfo {author} {\bibfnamefont {T.}~\bibnamefont
  {Homer-Dixon}}, \bibinfo {author} {\bibfnamefont {S.}~\bibnamefont
  {Janzwood}}, \bibinfo {author} {\bibfnamefont {J.}~\bibnamefont
  {Rockstr{\"o}m}}, \bibinfo {author} {\bibfnamefont {O.}~\bibnamefont
  {Renn}},\ and\ \bibinfo {author} {\bibfnamefont {J.~F.}\ \bibnamefont
  {Donges}},\ }\bibfield  {title} {\bibinfo {title} {Global polycrisis: the
  causal mechanisms of crisis entanglement},\ }\href
  {https://doi.org/10.1017/sus.2024.1} {\bibfield  {journal} {\bibinfo
  {journal} {Global Sustainability}\ }\textbf {\bibinfo {volume} {7}},\
  \bibinfo {pages} {e6} (\bibinfo {year} {2024})}\BibitemShut {NoStop}%
\bibitem [{\citenamefont {Rockstr{\"o}m}\ \emph {et~al.}(2009)\citenamefont
  {Rockstr{\"o}m}, \citenamefont {Steffen}, \citenamefont {Noone},
  \citenamefont {Persson}, \citenamefont {Chapin}, \citenamefont {Lambin},
  \citenamefont {Lenton}, \citenamefont {Scheffer}, \citenamefont {Folke},
  \citenamefont {Schellnhuber}, \citenamefont {Nykvist}, \citenamefont
  {De~Wit}, \citenamefont {Hughes}, \citenamefont {van~der Leeuw},
  \citenamefont {Rodhe}, \citenamefont {S{\"o}rlin}, \citenamefont {Snyder},
  \citenamefont {Costanza}, \citenamefont {Svedin}, \citenamefont {Falkenmark},
  \citenamefont {Karlberg}, \citenamefont {Corell}, \citenamefont {Fabry},
  \citenamefont {Hansen}, \citenamefont {Walker}, \citenamefont {Liverman},
  \citenamefont {Richardson}, \citenamefont {Crutzen},\ and\ \citenamefont
  {Foley}}]{Rockstrom:2009}%
  \BibitemOpen
  \bibfield  {author} {\bibinfo {author} {\bibfnamefont {J.}~\bibnamefont
  {Rockstr{\"o}m}}, \bibinfo {author} {\bibfnamefont {W.}~\bibnamefont
  {Steffen}}, \bibinfo {author} {\bibfnamefont {K.}~\bibnamefont {Noone}},
  \bibinfo {author} {\bibfnamefont {A.}~\bibnamefont {Persson}}, \bibinfo
  {author} {\bibfnamefont {F.~S.}\ \bibnamefont {Chapin}, \bibfnamefont {III}},
  \bibinfo {author} {\bibfnamefont {E.}~\bibnamefont {Lambin}}, \bibinfo
  {author} {\bibfnamefont {T.~M.}\ \bibnamefont {Lenton}}, \bibinfo {author}
  {\bibfnamefont {M.}~\bibnamefont {Scheffer}}, \bibinfo {author}
  {\bibfnamefont {C.}~\bibnamefont {Folke}}, \bibinfo {author} {\bibfnamefont
  {H.}~\bibnamefont {Schellnhuber}}, \bibinfo {author} {\bibfnamefont
  {B.}~\bibnamefont {Nykvist}}, \bibinfo {author} {\bibfnamefont {C.~A.}\
  \bibnamefont {De~Wit}}, \bibinfo {author} {\bibfnamefont {T.}~\bibnamefont
  {Hughes}}, \bibinfo {author} {\bibfnamefont {S.}~\bibnamefont {van~der
  Leeuw}}, \bibinfo {author} {\bibfnamefont {H.}~\bibnamefont {Rodhe}},
  \bibinfo {author} {\bibfnamefont {S.}~\bibnamefont {S{\"o}rlin}}, \bibinfo
  {author} {\bibfnamefont {P.~K.}\ \bibnamefont {Snyder}}, \bibinfo {author}
  {\bibfnamefont {R.}~\bibnamefont {Costanza}}, \bibinfo {author}
  {\bibfnamefont {U.}~\bibnamefont {Svedin}}, \bibinfo {author} {\bibfnamefont
  {M.}~\bibnamefont {Falkenmark}}, \bibinfo {author} {\bibfnamefont
  {L.}~\bibnamefont {Karlberg}}, \bibinfo {author} {\bibfnamefont {R.~W.}\
  \bibnamefont {Corell}}, \bibinfo {author} {\bibfnamefont {V.~J.}\
  \bibnamefont {Fabry}}, \bibinfo {author} {\bibfnamefont {J.}~\bibnamefont
  {Hansen}}, \bibinfo {author} {\bibfnamefont {B.}~\bibnamefont {Walker}},
  \bibinfo {author} {\bibfnamefont {D.}~\bibnamefont {Liverman}}, \bibinfo
  {author} {\bibfnamefont {K.}~\bibnamefont {Richardson}}, \bibinfo {author}
  {\bibfnamefont {P.}~\bibnamefont {Crutzen}},\ and\ \bibinfo {author}
  {\bibfnamefont {J.}~\bibnamefont {Foley}},\ }\bibfield  {title} {\bibinfo
  {title} {Planetary boundaries: Exploring the safe operating space for
  humanity},\ }\href@noop {} {\bibfield  {journal} {\bibinfo  {journal}
  {Ecology and Society}\ }\textbf {\bibinfo {volume} {14}},\ \bibinfo {pages}
  {32} (\bibinfo {year} {2009})}\BibitemShut {NoStop}%
\bibitem [{\citenamefont {Steffen}\ \emph {et~al.}(2015)\citenamefont
  {Steffen}, \citenamefont {Richardson}, \citenamefont {Rockstr{\"o}m},
  \citenamefont {Cornell}, \citenamefont {Fetzer}, \citenamefont {Bennett},
  \citenamefont {Biggs}, \citenamefont {Carpenter}, \citenamefont {de~Vries},
  \citenamefont {de~Wit}, \citenamefont {Folke}, \citenamefont {Gerten},
  \citenamefont {Heinke}, \citenamefont {Mace}, \citenamefont {Persson},
  \citenamefont {Ramanathan}, \citenamefont {Reyers},\ and\ \citenamefont
  {S{\"o}rlin}}]{Steffen:2015a}%
  \BibitemOpen
  \bibfield  {author} {\bibinfo {author} {\bibfnamefont {W.}~\bibnamefont
  {Steffen}}, \bibinfo {author} {\bibfnamefont {K.}~\bibnamefont {Richardson}},
  \bibinfo {author} {\bibfnamefont {J.}~\bibnamefont {Rockstr{\"o}m}}, \bibinfo
  {author} {\bibfnamefont {S.~E.}\ \bibnamefont {Cornell}}, \bibinfo {author}
  {\bibfnamefont {I.}~\bibnamefont {Fetzer}}, \bibinfo {author} {\bibfnamefont
  {E.~M.}\ \bibnamefont {Bennett}}, \bibinfo {author} {\bibfnamefont
  {R.}~\bibnamefont {Biggs}}, \bibinfo {author} {\bibfnamefont {S.~R.}\
  \bibnamefont {Carpenter}}, \bibinfo {author} {\bibfnamefont {W.}~\bibnamefont
  {de~Vries}}, \bibinfo {author} {\bibfnamefont {C.~A.}\ \bibnamefont
  {de~Wit}}, \bibinfo {author} {\bibfnamefont {C.}~\bibnamefont {Folke}},
  \bibinfo {author} {\bibfnamefont {D.}~\bibnamefont {Gerten}}, \bibinfo
  {author} {\bibfnamefont {J.}~\bibnamefont {Heinke}}, \bibinfo {author}
  {\bibfnamefont {G.~M.}\ \bibnamefont {Mace}}, \bibinfo {author}
  {\bibfnamefont {L.~M.}\ \bibnamefont {Persson}}, \bibinfo {author}
  {\bibfnamefont {V.}~\bibnamefont {Ramanathan}}, \bibinfo {author}
  {\bibfnamefont {B.}~\bibnamefont {Reyers}},\ and\ \bibinfo {author}
  {\bibfnamefont {S.}~\bibnamefont {S{\"o}rlin}},\ }\bibfield  {title}
  {\bibinfo {title} {Planetary {B}oundaries: Guiding human development on a
  changing planet},\ }\href {https://doi.org/10.1126/science.1259855}
  {\bibfield  {journal} {\bibinfo  {journal} {Science}\ }\textbf {\bibinfo
  {volume} {347}},\ \bibinfo {pages} {1259855} (\bibinfo {year}
  {2015})}\BibitemShut {NoStop}%
\bibitem [{\citenamefont {Lenton}\ \emph {et~al.}(2008)\citenamefont {Lenton},
  \citenamefont {Held}, \citenamefont {Kriegler}, \citenamefont {Hall},
  \citenamefont {Lucht}, \citenamefont {Rahmstorf},\ and\ \citenamefont
  {Schellnhuber}}]{Lenton:2008}%
  \BibitemOpen
  \bibfield  {author} {\bibinfo {author} {\bibfnamefont {T.~M.}\ \bibnamefont
  {Lenton}}, \bibinfo {author} {\bibfnamefont {H.}~\bibnamefont {Held}},
  \bibinfo {author} {\bibfnamefont {E.}~\bibnamefont {Kriegler}}, \bibinfo
  {author} {\bibfnamefont {J.~W.}\ \bibnamefont {Hall}}, \bibinfo {author}
  {\bibfnamefont {W.}~\bibnamefont {Lucht}}, \bibinfo {author} {\bibfnamefont
  {S.}~\bibnamefont {Rahmstorf}},\ and\ \bibinfo {author} {\bibfnamefont
  {H.~J.}\ \bibnamefont {Schellnhuber}},\ }\bibfield  {title} {\bibinfo {title}
  {Tipping elements in the earth's climate system},\ }\href@noop {} {\bibfield
  {journal} {\bibinfo  {journal} {Proceedings of the National Academy of
  Sciences}\ }\textbf {\bibinfo {volume} {105}},\ \bibinfo {pages} {1786}
  (\bibinfo {year} {2008})}\BibitemShut {NoStop}%
\bibitem [{\citenamefont {Smith}\ \emph {et~al.}(2022)\citenamefont {Smith},
  \citenamefont {Traxl},\ and\ \citenamefont {Boers}}]{Smith:2022}%
  \BibitemOpen
  \bibfield  {author} {\bibinfo {author} {\bibfnamefont {T.}~\bibnamefont
  {Smith}}, \bibinfo {author} {\bibfnamefont {D.}~\bibnamefont {Traxl}},\ and\
  \bibinfo {author} {\bibfnamefont {N.}~\bibnamefont {Boers}},\ }\bibfield
  {title} {\bibinfo {title} {Empirical evidence for recent global shifts in
  vegetation resilience},\ }\href@noop {} {\bibfield  {journal} {\bibinfo
  {journal} {Nature Climate Change}\ }\textbf {\bibinfo {volume} {12}},\
  \bibinfo {pages} {477} (\bibinfo {year} {2022})}\BibitemShut {NoStop}%
\bibitem [{\citenamefont {Haddad}\ \emph {et~al.}(2015)\citenamefont {Haddad},
  \citenamefont {Brudvig}, \citenamefont {Clobert}, \citenamefont {Davies},
  \citenamefont {Gonzalez}, \citenamefont {Holt}, \citenamefont {Lovejoy},
  \citenamefont {Sexton}, \citenamefont {Austin}, \citenamefont {Collins},
  \citenamefont {Cook}, \citenamefont {Damschen}, \citenamefont {Ewers},
  \citenamefont {Foster}, \citenamefont {Jenkins}, \citenamefont {King},
  \citenamefont {Laurance}, \citenamefont {Levey}, \citenamefont {Margules},
  \citenamefont {Melbourne}, \citenamefont {Nicholls}, \citenamefont {Orrock},
  \citenamefont {Song},\ and\ \citenamefont {Townshend}}]{Haddad:2015}%
  \BibitemOpen
  \bibfield  {author} {\bibinfo {author} {\bibfnamefont {N.~M.}\ \bibnamefont
  {Haddad}}, \bibinfo {author} {\bibfnamefont {L.~A.}\ \bibnamefont {Brudvig}},
  \bibinfo {author} {\bibfnamefont {J.}~\bibnamefont {Clobert}}, \bibinfo
  {author} {\bibfnamefont {K.~F.}\ \bibnamefont {Davies}}, \bibinfo {author}
  {\bibfnamefont {A.}~\bibnamefont {Gonzalez}}, \bibinfo {author}
  {\bibfnamefont {R.~D.}\ \bibnamefont {Holt}}, \bibinfo {author}
  {\bibfnamefont {T.~E.}\ \bibnamefont {Lovejoy}}, \bibinfo {author}
  {\bibfnamefont {J.~O.}\ \bibnamefont {Sexton}}, \bibinfo {author}
  {\bibfnamefont {M.~P.}\ \bibnamefont {Austin}}, \bibinfo {author}
  {\bibfnamefont {C.~D.}\ \bibnamefont {Collins}}, \bibinfo {author}
  {\bibfnamefont {W.~M.}\ \bibnamefont {Cook}}, \bibinfo {author}
  {\bibfnamefont {E.~I.}\ \bibnamefont {Damschen}}, \bibinfo {author}
  {\bibfnamefont {R.~M.}\ \bibnamefont {Ewers}}, \bibinfo {author}
  {\bibfnamefont {B.~L.}\ \bibnamefont {Foster}}, \bibinfo {author}
  {\bibfnamefont {C.~N.}\ \bibnamefont {Jenkins}}, \bibinfo {author}
  {\bibfnamefont {A.~J.}\ \bibnamefont {King}}, \bibinfo {author}
  {\bibfnamefont {W.~F.}\ \bibnamefont {Laurance}}, \bibinfo {author}
  {\bibfnamefont {D.~J.}\ \bibnamefont {Levey}}, \bibinfo {author}
  {\bibfnamefont {C.~R.}\ \bibnamefont {Margules}}, \bibinfo {author}
  {\bibfnamefont {B.~A.}\ \bibnamefont {Melbourne}}, \bibinfo {author}
  {\bibfnamefont {A.~O.}\ \bibnamefont {Nicholls}}, \bibinfo {author}
  {\bibfnamefont {J.~L.}\ \bibnamefont {Orrock}}, \bibinfo {author}
  {\bibfnamefont {D.-X.}\ \bibnamefont {Song}},\ and\ \bibinfo {author}
  {\bibfnamefont {J.~R.}\ \bibnamefont {Townshend}},\ }\bibfield  {title}
  {\bibinfo {title} {Habitat fragmentation and its lasting impact on {E}arth's
  ecosystems},\ }\href {https://doi.org/10.1126/sciadv.1500052} {\bibfield
  {journal} {\bibinfo  {journal} {Science Advances}\ }\textbf {\bibinfo
  {volume} {1}},\ \bibinfo {pages} {e1500052} (\bibinfo {year}
  {2015})}\BibitemShut {NoStop}%
\bibitem [{\citenamefont {McKay}\ \emph {et~al.}(2022)\citenamefont {McKay},
  \citenamefont {Staal}, \citenamefont {Abrams}, \citenamefont {Winkelmann},
  \citenamefont {Sakschewski}, \citenamefont {Loriani}, \citenamefont {Fetzer},
  \citenamefont {Cornell}, \citenamefont {Rockstr{\"o}m},\ and\ \citenamefont
  {Lenton}}]{McKay:2022}%
  \BibitemOpen
  \bibfield  {author} {\bibinfo {author} {\bibfnamefont {D.~A.}\ \bibnamefont
  {McKay}}, \bibinfo {author} {\bibfnamefont {A.}~\bibnamefont {Staal}},
  \bibinfo {author} {\bibfnamefont {J.~F.}\ \bibnamefont {Abrams}}, \bibinfo
  {author} {\bibfnamefont {R.}~\bibnamefont {Winkelmann}}, \bibinfo {author}
  {\bibfnamefont {B.}~\bibnamefont {Sakschewski}}, \bibinfo {author}
  {\bibfnamefont {S.}~\bibnamefont {Loriani}}, \bibinfo {author} {\bibfnamefont
  {I.}~\bibnamefont {Fetzer}}, \bibinfo {author} {\bibfnamefont {S.~E.}\
  \bibnamefont {Cornell}}, \bibinfo {author} {\bibfnamefont {J.}~\bibnamefont
  {Rockstr{\"o}m}},\ and\ \bibinfo {author} {\bibfnamefont {T.~M.}\
  \bibnamefont {Lenton}},\ }\bibfield  {title} {\bibinfo {title} {Exceeding
  1.5\textdegree {C} global warming could trigger multiple climate tipping
  points},\ }\href@noop {} {\bibfield  {journal} {\bibinfo  {journal}
  {Science}\ }\textbf {\bibinfo {volume} {377}},\ \bibinfo {pages} {eabn7950}
  (\bibinfo {year} {2022})}\BibitemShut {NoStop}%
\bibitem [{\citenamefont {Scheffer}\ \emph {et~al.}(2012)\citenamefont
  {Scheffer}, \citenamefont {Carpenter}, \citenamefont {Lenton}, \citenamefont
  {Bascompte}, \citenamefont {Brock}, \citenamefont {Dakos}, \citenamefont
  {van~de Koppel}, \citenamefont {van~de Leemput}, \citenamefont {Levin},
  \citenamefont {van Nes}, \citenamefont {Pascual},\ and\ \citenamefont
  {Vandermeer}}]{Scheffer:2012}%
  \BibitemOpen
  \bibfield  {author} {\bibinfo {author} {\bibfnamefont {M.}~\bibnamefont
  {Scheffer}}, \bibinfo {author} {\bibfnamefont {S.~R.}\ \bibnamefont
  {Carpenter}}, \bibinfo {author} {\bibfnamefont {T.~M.}\ \bibnamefont
  {Lenton}}, \bibinfo {author} {\bibfnamefont {J.}~\bibnamefont {Bascompte}},
  \bibinfo {author} {\bibfnamefont {W.}~\bibnamefont {Brock}}, \bibinfo
  {author} {\bibfnamefont {V.}~\bibnamefont {Dakos}}, \bibinfo {author}
  {\bibfnamefont {J.}~\bibnamefont {van~de Koppel}}, \bibinfo {author}
  {\bibfnamefont {I.~A.}\ \bibnamefont {van~de Leemput}}, \bibinfo {author}
  {\bibfnamefont {S.~A.}\ \bibnamefont {Levin}}, \bibinfo {author}
  {\bibfnamefont {E.~H.}\ \bibnamefont {van Nes}}, \bibinfo {author}
  {\bibfnamefont {M.}~\bibnamefont {Pascual}},\ and\ \bibinfo {author}
  {\bibfnamefont {J.}~\bibnamefont {Vandermeer}},\ }\bibfield  {title}
  {\bibinfo {title} {Anticipating critical transitions},\ }\href@noop {}
  {\bibfield  {journal} {\bibinfo  {journal} {Science}\ }\textbf {\bibinfo
  {volume} {338}},\ \bibinfo {pages} {344} (\bibinfo {year}
  {2012})}\BibitemShut {NoStop}%
\bibitem [{\citenamefont {Scheffer}\ \emph {et~al.}(2009)\citenamefont
  {Scheffer}, \citenamefont {Bascompte}, \citenamefont {Brock}, \citenamefont
  {Brovkin}, \citenamefont {Carpenter}, \citenamefont {Dakos}, \citenamefont
  {Held}, \citenamefont {van Nes}, \citenamefont {Rietkerk},\ and\
  \citenamefont {Sugihara}}]{Scheffer:2009}%
  \BibitemOpen
  \bibfield  {author} {\bibinfo {author} {\bibfnamefont {M.}~\bibnamefont
  {Scheffer}}, \bibinfo {author} {\bibfnamefont {J.}~\bibnamefont {Bascompte}},
  \bibinfo {author} {\bibfnamefont {W.~A.}\ \bibnamefont {Brock}}, \bibinfo
  {author} {\bibfnamefont {V.}~\bibnamefont {Brovkin}}, \bibinfo {author}
  {\bibfnamefont {S.~R.}\ \bibnamefont {Carpenter}}, \bibinfo {author}
  {\bibfnamefont {V.}~\bibnamefont {Dakos}}, \bibinfo {author} {\bibfnamefont
  {H.}~\bibnamefont {Held}}, \bibinfo {author} {\bibfnamefont {E.~H.}\
  \bibnamefont {van Nes}}, \bibinfo {author} {\bibfnamefont {M.}~\bibnamefont
  {Rietkerk}},\ and\ \bibinfo {author} {\bibfnamefont {G.}~\bibnamefont
  {Sugihara}},\ }\bibfield  {title} {\bibinfo {title} {Early-warning signals
  for critical transitions},\ }\href {https://doi.org/10.1038/nature08227}
  {\bibfield  {journal} {\bibinfo  {journal} {Nature}\ }\textbf {\bibinfo
  {volume} {461}},\ \bibinfo {pages} {53} (\bibinfo {year} {2009})}\BibitemShut
  {NoStop}%
\bibitem [{\citenamefont {Caesar}\ \emph {et~al.}(2024)\citenamefont {Caesar},
  \citenamefont {Sakschewski}, \citenamefont {Andersen}, \citenamefont
  {Beringer}, \citenamefont {Braun}, \citenamefont {Dennis}, \citenamefont
  {Gerten}, \citenamefont {Heilemann}, \citenamefont {Kaiser}, \citenamefont
  {Kitzmann}, \citenamefont {Loriani}, \citenamefont {Lucht}, \citenamefont
  {Ludescher}, \citenamefont {Martin}, \citenamefont {Mathesius}, \citenamefont
  {Paolucci}, \citenamefont {te~Wierik},\ and\ \citenamefont
  {Rockstr\"om}}]{Caeser:2024}%
  \BibitemOpen
  \bibfield  {author} {\bibinfo {author} {\bibfnamefont {L.}~\bibnamefont
  {Caesar}}, \bibinfo {author} {\bibfnamefont {B.}~\bibnamefont {Sakschewski}},
  \bibinfo {author} {\bibfnamefont {L.~S.}\ \bibnamefont {Andersen}}, \bibinfo
  {author} {\bibfnamefont {T.}~\bibnamefont {Beringer}}, \bibinfo {author}
  {\bibfnamefont {J.}~\bibnamefont {Braun}}, \bibinfo {author} {\bibfnamefont
  {D.}~\bibnamefont {Dennis}}, \bibinfo {author} {\bibfnamefont
  {D.}~\bibnamefont {Gerten}}, \bibinfo {author} {\bibfnamefont
  {A.}~\bibnamefont {Heilemann}}, \bibinfo {author} {\bibfnamefont
  {J.}~\bibnamefont {Kaiser}}, \bibinfo {author} {\bibfnamefont
  {N.}~\bibnamefont {Kitzmann}}, \bibinfo {author} {\bibfnamefont
  {S.}~\bibnamefont {Loriani}}, \bibinfo {author} {\bibfnamefont
  {W.}~\bibnamefont {Lucht}}, \bibinfo {author} {\bibfnamefont
  {J.}~\bibnamefont {Ludescher}}, \bibinfo {author} {\bibfnamefont
  {M.}~\bibnamefont {Martin}}, \bibinfo {author} {\bibfnamefont
  {S.}~\bibnamefont {Mathesius}}, \bibinfo {author} {\bibfnamefont
  {A.}~\bibnamefont {Paolucci}}, \bibinfo {author} {\bibfnamefont
  {S.}~\bibnamefont {te~Wierik}},\ and\ \bibinfo {author} {\bibfnamefont
  {J.}~\bibnamefont {Rockstr\"om}},\ }\href@noop {} {\emph {\bibinfo {title}
  {Planetary Health Check Report 2024}}}\ (\bibinfo  {publisher} {Potsdam
  Institute for Climate Impact Research},\ \bibinfo {year} {2024})\BibitemShut
  {NoStop}%
\bibitem [{\citenamefont {Bertolami}\ and\ \citenamefont
  {Francisco}(2018)}]{BertolamiFrancisco:2018}%
  \BibitemOpen
  \bibfield  {author} {\bibinfo {author} {\bibfnamefont {O.}~\bibnamefont
  {Bertolami}}\ and\ \bibinfo {author} {\bibfnamefont {F.}~\bibnamefont
  {Francisco}},\ }\bibfield  {title} {\bibinfo {title} {A physical framework
  for the earth system, anthropocene equation and the great acceleration},\
  }\href@noop {} {\bibfield  {journal} {\bibinfo  {journal} {Global and
  Planetary Change}\ }\textbf {\bibinfo {volume} {169}},\ \bibinfo {pages} {66}
  (\bibinfo {year} {2018})}\BibitemShut {NoStop}%
\bibitem [{\citenamefont {Bertolami}\ and\ \citenamefont
  {Francisco}(2019)}]{BertolamiFrancisco:2019}%
  \BibitemOpen
  \bibfield  {author} {\bibinfo {author} {\bibfnamefont {O.}~\bibnamefont
  {Bertolami}}\ and\ \bibinfo {author} {\bibfnamefont {F.}~\bibnamefont
  {Francisco}},\ }\bibfield  {title} {\bibinfo {title} {A phase space
  description of the earth system in the anthropocene},\ }\href@noop {}
  {\bibfield  {journal} {\bibinfo  {journal} {EPL (Europhysics Letters)}\
  }\textbf {\bibinfo {volume} {127}},\ \bibinfo {pages} {59001} (\bibinfo
  {year} {2019})}\BibitemShut {NoStop}%
\bibitem [{\citenamefont {Barbosa}\ \emph {et~al.}(2020)\citenamefont
  {Barbosa}, \citenamefont {Bertolami},\ and\ \citenamefont
  {Francisco}}]{Barbosa:2020}%
  \BibitemOpen
  \bibfield  {author} {\bibinfo {author} {\bibfnamefont {M.}~\bibnamefont
  {Barbosa}}, \bibinfo {author} {\bibfnamefont {O.}~\bibnamefont {Bertolami}},\
  and\ \bibinfo {author} {\bibfnamefont {F.}~\bibnamefont {Francisco}},\
  }\bibfield  {title} {\bibinfo {title} {Towards a physically motivated
  planetary accounting framework},\ }\href
  {https://doi.org/10.1177/2053019620909659} {\bibfield  {journal} {\bibinfo
  {journal} {The Anthropocene Review}\ }\textbf {\bibinfo {volume} {7}},\
  \bibinfo {pages} {191} (\bibinfo {year} {2020})},\ \Eprint
  {https://arxiv.org/abs/https://doi.org/10.1177/2053019620909659}
  {https://doi.org/10.1177/2053019620909659} \BibitemShut {NoStop}%
\bibitem [{\citenamefont {Bertolami}\ and\ \citenamefont
  {Nystr\"om}(2025)}]{BertolamiNystrom:2025}%
  \BibitemOpen
  \bibfield  {author} {\bibinfo {author} {\bibfnamefont {O.}~\bibnamefont
  {Bertolami}}\ and\ \bibinfo {author} {\bibfnamefont {M.}~\bibnamefont
  {Nystr\"om}},\ }\bibfield  {title} {\bibinfo {title} {Setting up the physical
  principles of resilience in a model of the earth system},\ }\href@noop {}
  {\bibfield  {journal} {\bibinfo  {journal} {submitted for publication}\ }
  (\bibinfo {year} {2025})}\BibitemShut {NoStop}%
\bibitem [{\citenamefont {Steffen}\ \emph {et~al.}(2018)\citenamefont
  {Steffen}, \citenamefont {Rockstr{\"o}m}, \citenamefont {Richardson},
  \citenamefont {Lenton}, \citenamefont {Folke}, \citenamefont {Liverman},
  \citenamefont {Summerhayes}, \citenamefont {Barnosky}, \citenamefont
  {Cornell}, \citenamefont {Crucifix}, \citenamefont {Donges}, \citenamefont
  {Fetzer}, \citenamefont {Lade}, \citenamefont {Scheffer}, \citenamefont
  {Winkelmann},\ and\ \citenamefont {Schellnhuber}}]{Steffen:2018}%
  \BibitemOpen
  \bibfield  {author} {\bibinfo {author} {\bibfnamefont {W.}~\bibnamefont
  {Steffen}}, \bibinfo {author} {\bibfnamefont {J.}~\bibnamefont
  {Rockstr{\"o}m}}, \bibinfo {author} {\bibfnamefont {K.}~\bibnamefont
  {Richardson}}, \bibinfo {author} {\bibfnamefont {T.~M.}\ \bibnamefont
  {Lenton}}, \bibinfo {author} {\bibfnamefont {C.}~\bibnamefont {Folke}},
  \bibinfo {author} {\bibfnamefont {D.}~\bibnamefont {Liverman}}, \bibinfo
  {author} {\bibfnamefont {C.~P.}\ \bibnamefont {Summerhayes}}, \bibinfo
  {author} {\bibfnamefont {A.~D.}\ \bibnamefont {Barnosky}}, \bibinfo {author}
  {\bibfnamefont {S.~E.}\ \bibnamefont {Cornell}}, \bibinfo {author}
  {\bibfnamefont {M.}~\bibnamefont {Crucifix}}, \bibinfo {author}
  {\bibfnamefont {J.~F.}\ \bibnamefont {Donges}}, \bibinfo {author}
  {\bibfnamefont {I.}~\bibnamefont {Fetzer}}, \bibinfo {author} {\bibfnamefont
  {S.~J.}\ \bibnamefont {Lade}}, \bibinfo {author} {\bibfnamefont
  {M.}~\bibnamefont {Scheffer}}, \bibinfo {author} {\bibfnamefont
  {R.}~\bibnamefont {Winkelmann}},\ and\ \bibinfo {author} {\bibfnamefont
  {H.~J.}\ \bibnamefont {Schellnhuber}},\ }\bibfield  {title} {\bibinfo {title}
  {Trajectories of the earth system in the anthropocene},\ }\href
  {https://doi.org/10.1073/pnas.1810141115} {\bibfield  {journal} {\bibinfo
  {journal} {Proceedings of the National Academy of Sciences}\ }\textbf
  {\bibinfo {volume} {115}},\ \bibinfo {pages} {8252} (\bibinfo {year}
  {2018})}\BibitemShut {NoStop}%
\bibitem [{\citenamefont {Bernardini}\ \emph {et~al.}(2025)\citenamefont
  {Bernardini}, \citenamefont {Bertolami},\ and\ \citenamefont
  {Francisco}}]{Bernardini:2025}%
  \BibitemOpen
  \bibfield  {author} {\bibinfo {author} {\bibfnamefont {A.~E.}\ \bibnamefont
  {Bernardini}}, \bibinfo {author} {\bibfnamefont {O.}~\bibnamefont
  {Bertolami}},\ and\ \bibinfo {author} {\bibfnamefont {F.}~\bibnamefont
  {Francisco}},\ }\bibfield  {title} {\bibinfo {title} {Chaotic behaviour of
  the earth system in the anthropocene},\ }\href@noop {} {\bibfield  {journal}
  {\bibinfo  {journal} {Evolving Earth}\ }\textbf {\bibinfo {volume} {3}},\
  \bibinfo {pages} {100060} (\bibinfo {year} {2025})}\BibitemShut {NoStop}%
\bibitem [{\citenamefont {S{\o}gaard~J{\o}rgensen}\ \emph
  {et~al.}(2024)\citenamefont {S{\o}gaard~J{\o}rgensen}, \citenamefont
  {Jansen}, \citenamefont {Avila~Ortega}, \citenamefont {Wang-Erlandsson},
  \citenamefont {Donges}, \citenamefont {\"Osterblom}, \citenamefont {Olsson},
  \citenamefont {Nystr\"om}, \citenamefont {Lade}, \citenamefont {Hahn},
  \citenamefont {Folke}, \citenamefont {Peterson},\ and\ \citenamefont
  {Cr\'epin}}]{Jorgensen:2024}%
  \BibitemOpen
  \bibfield  {author} {\bibinfo {author} {\bibfnamefont {P.}~\bibnamefont
  {S{\o}gaard~J{\o}rgensen}}, \bibinfo {author} {\bibfnamefont {R.~E.~V.}\
  \bibnamefont {Jansen}}, \bibinfo {author} {\bibfnamefont {D.~I.}\
  \bibnamefont {Avila~Ortega}}, \bibinfo {author} {\bibfnamefont
  {L.}~\bibnamefont {Wang-Erlandsson}}, \bibinfo {author} {\bibfnamefont
  {J.~F.}\ \bibnamefont {Donges}}, \bibinfo {author} {\bibfnamefont
  {H.}~\bibnamefont {\"Osterblom}}, \bibinfo {author} {\bibfnamefont
  {P.}~\bibnamefont {Olsson}}, \bibinfo {author} {\bibfnamefont
  {M.}~\bibnamefont {Nystr\"om}}, \bibinfo {author} {\bibfnamefont {S.~J.}\
  \bibnamefont {Lade}}, \bibinfo {author} {\bibfnamefont {T.}~\bibnamefont
  {Hahn}}, \bibinfo {author} {\bibfnamefont {C.}~\bibnamefont {Folke}},
  \bibinfo {author} {\bibfnamefont {G.~D.}\ \bibnamefont {Peterson}},\ and\
  \bibinfo {author} {\bibfnamefont {A.-S.}\ \bibnamefont {Cr\'epin}},\
  }\bibfield  {title} {\bibinfo {title} {Evolution of the polycrisis:
  Anthropocene traps that challenge global sustainability},\ }\href@noop {}
  {\bibfield  {journal} {\bibinfo  {journal} {Phil. Trans. R. Soc.}\ }
  (\bibinfo {year} {2024})}\BibitemShut {NoStop}%
\bibitem [{\citenamefont {Ising}(1925)}]{Ising:1925}%
  \BibitemOpen
  \bibfield  {author} {\bibinfo {author} {\bibfnamefont {E.}~\bibnamefont
  {Ising}},\ }\bibfield  {title} {\bibinfo {title} {Contribution to the theory
  of ferromagnetism},\ }\href@noop {} {\bibfield  {journal} {\bibinfo
  {journal} {Zeitschrift für Physik}\ }\textbf {\bibinfo {volume} {XXXI}}
  (\bibinfo {year} {1925})}\BibitemShut {NoStop}%
\bibitem [{\citenamefont {Halperin}(1982)}]{Halperin:1982}%
  \BibitemOpen
  \bibfield  {author} {\bibinfo {author} {\bibfnamefont {B.}~\bibnamefont
  {Halperin}},\ }\bibfield  {title} {\bibinfo {title} {Quantized hall
  conductance, current-carrying edge states, and the existence of extended
  states in a two-dimensional disordered potential},\ }\href
  {https://doi.org/10.1103/PhysRevLett.49.405} {\bibfield  {journal} {\bibinfo
  {journal} {Physical Review Letters}\ }\textbf {\bibinfo {volume} {49}},\
  \bibinfo {pages} {405} (\bibinfo {year} {1982})}\BibitemShut {NoStop}%
\bibitem [{\citenamefont {Haldane}(1983)}]{Haldane:1983}%
  \BibitemOpen
  \bibfield  {author} {\bibinfo {author} {\bibfnamefont {F.}~\bibnamefont
  {Haldane}},\ }\bibfield  {title} {\bibinfo {title} {Nonlinear field theory of
  large-spin heisenberg antiferromagnets: Semiclassically quantized solitons of
  the one-dimensional easy-axis néel state},\ }\href
  {https://doi.org/10.1103/PhysRevLett.50.1153} {\bibfield  {journal} {\bibinfo
   {journal} {Physical Review Letters}\ }\textbf {\bibinfo {volume} {50}},\
  \bibinfo {pages} {1153} (\bibinfo {year} {1983})}\BibitemShut {NoStop}%
\bibitem [{\citenamefont {Banks}\ \emph {et~al.}(1997)\citenamefont {Banks},
  \citenamefont {Fischler}, \citenamefont {Shenker},\ and\ \citenamefont
  {Susskind}}]{BFSS:1997}%
  \BibitemOpen
  \bibfield  {author} {\bibinfo {author} {\bibfnamefont {T.}~\bibnamefont
  {Banks}}, \bibinfo {author} {\bibfnamefont {W.}~\bibnamefont {Fischler}},
  \bibinfo {author} {\bibfnamefont {S.~H.}\ \bibnamefont {Shenker}},\ and\
  \bibinfo {author} {\bibfnamefont {L.}~\bibnamefont {Susskind}},\ }\bibfield
  {title} {\bibinfo {title} {M theory as a matrix model: A conjecture},\ }\href
  {https://doi.org/10.1103/PhysRevD.55.5112S} {\bibfield  {journal} {\bibinfo
  {journal} {Physical Review D}\ }\textbf {\bibinfo {volume} {55}},\ \bibinfo
  {pages} {5112} (\bibinfo {year} {1997})}\BibitemShut {NoStop}%
\bibitem [{\citenamefont {Verhulst}(1838)}]{Verhulst:1838}%
  \BibitemOpen
  \bibfield  {author} {\bibinfo {author} {\bibfnamefont {P.-F.}\ \bibnamefont
  {Verhulst}},\ }\bibfield  {title} {\bibinfo {title} {Notice sur la loi que la
  population poursuit dans son accroissement},\ }\href@noop {} {\bibfield
  {journal} {\bibinfo  {journal} {Correspondance Math{\'e}matique et Physique}\
  }\textbf {\bibinfo {volume} {10}},\ \bibinfo {pages} {113} (\bibinfo {year}
  {1838})}\BibitemShut {NoStop}%
\bibitem [{\citenamefont {Baker}\ and\ \citenamefont
  {Graves-Morris}(1996)}]{Baker:1996}%
  \BibitemOpen
  \bibfield  {author} {\bibinfo {author} {\bibfnamefont {G.~A.}\ \bibnamefont
  {Baker}}\ and\ \bibinfo {author} {\bibfnamefont {P.}~\bibnamefont
  {Graves-Morris}},\ }\href@noop {} {\emph {\bibinfo {title} {Pad{\'e}
  Approximants}}},\ \bibinfo {edition} {2nd}\ ed.,\ Encyclopedia of Mathematics
  and its Applications\ (\bibinfo  {publisher} {Cambridge University Press},\
  \bibinfo {year} {1996})\BibitemShut {NoStop}%
\bibitem [{\citenamefont {Wanner}\ \emph {et~al.}(1978)\citenamefont {Wanner},
  \citenamefont {Hairer},\ and\ \citenamefont {N{\o}rsett}}]{Wanner:1978}%
  \BibitemOpen
  \bibfield  {author} {\bibinfo {author} {\bibfnamefont {G.}~\bibnamefont
  {Wanner}}, \bibinfo {author} {\bibfnamefont {E.}~\bibnamefont {Hairer}},\
  and\ \bibinfo {author} {\bibfnamefont {S.}~\bibnamefont {N{\o}rsett}},\
  }\bibfield  {title} {\bibinfo {title} {Order stars and stability theorems},\
  }\href {https://doi.org/10.1007/BF01932026} {\bibfield  {journal} {\bibinfo
  {journal} {BIT}\ }\textbf {\bibinfo {volume} {18}},\ \bibinfo {pages} {475}
  (\bibinfo {year} {1978})}\BibitemShut {NoStop}%
\bibitem [{Note1()}]{Note1}%
  \BibitemOpen
  \bibinfo {note} {Given the SO(n) symmetry of the interaction, the emerging
  space structure of the Syndromes correspond to the one of k-forms in $l$
  dimensions \cite {Schutz:1980}.}\BibitemShut {Stop}%
\end{thebibliography}%

\end{document}